\documentclass[sigconf]{acmart}

\usepackage{booktabs} 

\usepackage[ruled]{algorithm2e} 
\usepackage{graphicx}
\usepackage{textcomp}
\usepackage{color,soul}
\usepackage{bm}
\usepackage{multirow}
\usepackage{subcaption}
\usepackage{balance}
\usepackage{graphicx}  

\usepackage{capt-of}
\usepackage{colortbl}
\usepackage{arydshln}
\usepackage{longtable,booktabs}
\usepackage{enumitem}
\usepackage{tcolorbox}

\definecolor{green}{rgb}{0.0, 0.65, 0.31}
\definecolor{bleudefrance}{rgb}{0.19, 0.55, 0.91}
\definecolor{ceruleanblue}{rgb}{0.16, 0.32, 0.75}
\definecolor{grey}{HTML}{969696}
\definecolor{violet}{HTML}{8856a7}
\definecolor{dgrey}{HTML}{01665e}
\definecolor{lgrey}{HTML}{5ab4ac}
\definecolor{dgreen}{HTML}{005a32}
\definecolor{purple}{HTML}{ae017e}
\definecolor{NewBlue}{HTML}{253494}
\definecolor{lblue}{HTML}{decbe4}

\definecolor{editCol}{HTML}{000000}
\definecolor{maskCol}{HTML}{c51b7d}
\definecolor{lrColor}{HTML}{8856a7}
\definecolor{trColor}{HTML}{d01c8b}
\definecolor{ctColor}{HTML}{4dac26}
\definecolor{brickred}{HTML}{f03b20}
\definecolor{improveCol}{HTML}{253494}
\definecolor{worsenCol}{HTML}{d7191c}
\definecolor{lgreen}{HTML}{e0f3db}
\definecolor{dpink}{HTML}{CD1076}
\definecolor{pink}{HTML}{FED2D2}
\definecolor{soothinggreen}{HTML}{4dac26}
\definecolor{darkred}{HTML}{8B0000}

\definecolor{dblue}{HTML}{104E8B}
\definecolor{violet}{HTML}{8A2BE2}
\definecolor{mscolor}{HTML}{01665e}
\definecolor{nmscolor}{HTML}{d8b365}
\definecolor{deepgrey}{HTML}{525252}
\definecolor{dslate}{HTML}{2F4F4F}
\definecolor{dolive}{HTML}{556B2F}
\definecolor{teal}{HTML}{388E8E}
\definecolor{mscolor}{HTML}{01665e}
\definecolor{nmscolor}{HTML}{d8b365}

\definecolor{srcolor}{HTML}{e34a33}
\definecolor{smcolor}{HTML}{253494}
\definecolor{srsmcolor}{HTML}{7fcdbb}
\definecolor{bothcolor}{HTML}{fe9929}
\definecolor{onecolor}{HTML}{018571}

\definecolor{beet}{HTML}{8E388E}

\newcommand*{\textlabel}[2]{%
  \edef\@currentlabel{#1}
  \phantomsection
  #1\label{#2}
}

\definecolor{yellowHL}{HTML}{ffffb3}
\definecolor{blueHL}{HTML}{bebada}
\definecolor{greenHL}{HTML}{8dd3c7}
\definecolor{orangeHL}{HTML}{fb8072}

\newcommand{\SBSIE}[1]{\textsf{ESI}}
\newcommand{\SBJFP}[1]{\textsf{PJF}}
\newcommand{\SBWPS}[1]{\textsf{WPS}}
\newcommand{\SBNFT}[1]{\textsf{NFW}}

\colorlet{tableheadcolor}{gray!25} 
\colorlet{tablerowcolor}{gray!15} 
\colorlet{tablerowcolor2}{gray!12} 
\colorlet{tablerowcolor3}{gray!25} 
\colorlet{tablerowcolor4}{gray!50} 

\newcommand{\rowcollight}{\rowcolor{tablerowcolor2}} %

\newcommand{\Tr}{\textsf{Treatment}}

\newcommand{\Ct}{\textsf{Control}}

\renewcommand{\textrightarrow}{$\rightarrow$}

\colorlet{tableheadcolor}{gray!25} 
\colorlet{tablerowcolor}{gray!5} 

\definecolor{neutralCol}{HTML}{dd1c77}
\definecolor{neutralGreen}{HTML}{31a354}

\definecolor{bleudefrance}{rgb}{0.19, 0.55, 0.91}  
\definecolor{AfTrColor}{HTML}{0868ac}  
\definecolor{BfTrColor}{HTML}{a8ddb5}  

\definecolor{AfCtColor}{HTML}{b10026}  
\definecolor{BfCtColor}{HTML}{fd8d3c}

\graphicspath{ {figures/} }

\newif{\ifhidecomments}
 \hidecommentsfalse 
\ifhidecomments
       \newcommand{\yunhao}[1]{}
       \newcommand{\barbara}[1]{}
\newcommand{\koustuv}[1]{}
    \newcommand{\talayeh}[1]{}
\else
   \newcommand{\yunhao}[1]{\textbf{\sffamily{\textcolor{NewBlue}{[#1 -- Yunhao]}}}}
   \newcommand{\barbara}[1]{\textbf{\sffamily{\textcolor{violet}{[#1 -- Barbara]}}}}
\newcommand{\koustuv}[1]{\textbf{\sffamily{\textcolor{dpink}{[#1 -- Koustuv]}}}}
   \newcommand{\talayeh}[1]{\textbf{\sffamily{\textcolor{orange}{[#1 -- Talayeh]}}}}
  \fi
\AtBeginDocument{%
  \providecommand\BibTeX{{%
    \normalfont B\kern-0.5em{\scshape i\kern-0.25em b}\kern-0.8em\TeX}}}

\renewcommand\footnotetextcopyrightpermission[1]{} 
\acmConference[]{}{}{}
\setcopyright{none}
\begin{document}

\title[Social Media and Papageno Effect]{Mental Health Coping Stories on Social Media: \\ A Causal-Inference Study of Papageno Effect}

\author{Yunhao Yuan}
\email{yunhao.yuan@aalto.fi}
\affiliation{%
  \institution{Aalto University}
  \city{Espoo}
  \country{Finland}
  \postcode{02150}
}

\author{Koustuv Saha}
\email{koustuv.saha@gmail.com}
\affiliation{%
  \institution{Microsoft Research}
  \city{Montreal}
  \country{Canada}}


\author{Barbara Keller}
\email{barbara.keller@aalto.fi}
\affiliation{%
  \institution{Aalto University}
  \city{Espoo}
  \country{Finland}
  \postcode{02150}
}

\author{Erkki Tapio Isomets{\"a}}
\email{erkki.isometsa@hus.fi}
\affiliation{%
  \institution{University of Helsinki}
  \city{Espoo}
  \country{Finland}
  \postcode{02150}
}
\author{Talayeh Aledavood}
\email{talayeh.aledavood@aalto.fi}
\affiliation{%
  \institution{Aalto University}
  \city{Espoo}
  \country{Finland}
  \postcode{02150}
}





\begin{abstract}
The Papageno effect concerns how media can play a positive role in preventing and mitigating suicidal ideation and behaviors. With the increasing ubiquity and widespread use of social media, individuals often express and share lived experiences and struggles with mental health. However, there is a gap in our understanding about the existence and effectiveness of the Papageno effect in social media, which we study in this paper. In particular, we adopt a causal-inference framework to examine the impact of exposure to mental health coping stories on individuals on Twitter. We obtain a Twitter dataset with $\sim$2M posts by $\sim$10K individuals. We consider engaging with coping stories as the \Tr{} intervention, and adopt a stratified propensity score approach to find matched cohorts of \Tr{} and \Ct{} individuals. We measure the psychosocial shifts in affective, behavioral, and cognitive outcomes in longitudinal Twitter data before and after engaging with the coping stories. Our findings reveal that, engaging with coping stories leads to decreased stress and depression, and improved expressive writing, diversity, and interactivity. Our work discusses the practical and platform design implications in supporting mental wellbeing.
\end{abstract}

\ccsdesc[300]{Human-centered computing~Empirical studies in collaborative and social computing}
\ccsdesc[300]{Applied computing~Psychology}
\ccsdesc[300]{Human-centered computing~Social media}

\keywords{social media, mental health, suicidal ideation, natural language, causal inference, Papageno effect}


\maketitle

\section{Introduction}

According to a report from the World Health Organization ~\cite{who2019suicide},  globally, approximately 700,000 people fall victim to suicide each year. Suicide attempts and particularly committed suicides cause severe and tragic consequences among relatives and friends of the victims, as well as significant economic problems for society. Consequently, suicide has become a crucial global public health problem, and the World Health Organization has called for urgent action to reduce the suicide mortality rate.

While suicide is a combined outcome of multiple, interrelated factors, ranging from mental health issues to social factors, media can play an important role either in a harmful or beneficial direction. A considerable amount of literature~\cite{stack1987celebrities,fahey2018tracking,kumar2015detecting} has studied and re-confirmed the harmful effect of media, dubbed the ``Werther effect''~\cite{phillips1974influence}, describing a spike in suicides after a heavily publicized suicide. However, there is much less research about the beneficial effects of media, referred to as the ``Papageno effect'', 
describing a decrease in suicides after reporting alternatives to suicide.~\citeauthor{niederkrotenthaler2010role} explored the possible protective effect of media reporting about suicide~\cite{niederkrotenthaler2010role}. This study finds a decrease in suicides, if reports of suicide related content portray ways of overcoming suicidal ideation without narrating suicidal behaviors. This work provides important insights into the potential benefits of media that reports suicide related content with a focus on hope and recovery. Following this work, other studies provide evidence of the Papageno effect from fictional films~\cite{till2015determining}, suicide-educational websites~\cite{till2017beneficial}, and newspaper articles~\cite{arendt2016effects}. Given the prevalence and importance of social media, understanding more about the Papageno effect on social media can play a crucial role in decreasing suicide rates.

Studies of the Papageno effect commonly rely on self-reports, surveys, and publicly reported suicide statistics and only cover a small, selected group of people. People with suicidal ideation can face negative attitudes and stigmatization, which prevents many of them from seeking help~\cite{reynders2015help}. Additionally, the sensitive nature of suicide makes it challenging to collect data at scale on individuals who have suicidal ideations and conduct continuous  long-term follow-up studies.

The emergence of social media platforms, such as Twitter, Reddit, and TikTok, provides venues for people to not only connect with others, but also to express and share different aspects and life events of their personal lives~\cite{saha2021life}. 
Social media platforms provide a non-intrusive means to collect people's naturalistic data at scale. Research has leveraged social media data from different social media platforms to explore psychological and health issues using data from various domains such as drug misuse~\cite{garg2021detecting}, minority stress~\cite{yuan2022impact}, and mental health~\cite{coppersmith2018natural,saha2022social}. Social media platforms provide timely and relevant information on examining risk attributes longitudinally. The anonymous features of social media may reduce the biases found in research based on surveys and self-reported data. Consequently, social media data provide an unparalleled opportunity to study the Papageno effect in a broader population and its evolution over time.

In this paper, we leverage public data from Twitter, a popular social media site. We analyze longitudinal posts from Twitter users who reply to posts containing stories about coping with suicidal ideation. We examine psychosocial changes in affective, behavioral, and cognitive outcomes related to suicidal ideation. Specifically, we target the research questions of, \textit{whether the Papageno effect exists on Twitter and how we can quantify psychosocial changes of Twitter users before and after engaging with Twitter posts containing mental health coping stories.}

To achieve the research goals,  we collect 13,022 Twitter posts containing keyword phrases, which might indicate coping stories. We utilize a machine learning classifier from a previous study~\cite{metzler2021detecting} to automatically annotate the dataset and 3,077 Twitter posts are labeled as coping stories, which might result in the Papageno effect. Among them, we manually verify the accuracy of the classifier on a sampled coping story dataset. We collect data from two populations on Twitter: 787K posts from 2,468 individuals who reply to Twitter posts containing coping stories and 1.4M posts from 8,465 individuals in a control group. After applying stratified propensity score matching, we aggregate psychosocial outcomes as affective, behavioral, and cognitive outcomes and identify these with high significant effects. We observe that engaging with coping story posts on Twitter is linked to lower stress and depression, and higher expressive writing, diversity, and interactivity. 

\vspace{1em}
\noindent\textbf{Privacy and Ethics.}
Although we use public accessible data from Twitter, we are committed to protect the privacy of the data owners. We remove any information related to personal identity and paraphrase all quotations in this paper to avoid traceability. Given the sensitive nature of the topic of suicide and to avoid potential misuse, we adhere to Twitter data sharing standards and will only share the Twitter post IDs to other researchers. One of the authors is a certified psychiatrist. This helps us better understand our findings.





\section{Related Work}

\subsection{Media Effects on Suicidal Ideation}
The question of how media reports about suicide influence subsequent suicides has received considerable attention~\cite{niederkrotenthaler2019suicide}. For quite some time, studies focus on the negative impact of media portrayals on suicide and find a positive correlation between media coverage of suicidal behavior and suicidality~\cite{domaradzki2021werther}. These studies spark a debate about the possible preventive impact of media on suicide rates.

While one body of research highlights that increasing public understanding of mental health therapy may prevent suicide attempts~\cite{carmichael2019media}, others suggest that negative media coverage of suicides, such as the suicide victim's non-attractive features or the circumstances of the suicidal act, prevents imitative suicidal attempts~\cite{phillips1978airplane}. 
~\citeauthor{niederkrotenthaler2010role} discover that reports of people who considered suicide but afterward dealt with their problems constructively are linked to a short-term reduction in suicide rates in~\citeyear{niederkrotenthaler2010role}. This preventive effect is coined the ``Papageno effect''. Inspired by the Papageno effect,~\citeauthor{till2018effect}~\cite{till2018effect} conduct a randomized controlled trial to explore the beneficial impact of educative newspapers featuring suicide researchers in \citeyear{till2018effect}. They observe similar suicide-protective effects on both readers with, as well as readers without personal experience of suicide ideation. To further test the Papageno effect,~\citeauthor{niederkrotenthaler2022effects} conduct meta-analysis research and provide new evidence supporting the beneficial effect of media on individuals with suicidal ideation if the media narratives focus on hope and recovery from suicidal crises~\cite{niederkrotenthaler2022effects}.

So far, however, the Papageno effect on social media is understudied. Our work attempts to examine the psychosocial impacts of the Papageno effect on Twitter. We gather longitudinal social media data and compare multiple psychosocial outcomes of individuals engaging with coping story posts with a matched \Ct{} group.

\subsection{Mental Health and Psycholinguistics}
Although suicide is not an inevitable consequence of any psychiatric condition, research suggests a link between mental health and suicidal behaviors. According to a psychiatric autopsy study~\cite{cavanagh2003psychological}, more than 90\% of people who die by suicide suffer a mental disorder previous to their death. Patients who report anhedonia and sleeplessness with major anxiety symptoms, alcohol abuse, or emotional problems have the highest short-term risk for suicide~\cite{kleespies2000evidence}.

Studying suicidal ideation attracts researchers from different fields. Research suggests that psychological linguistic metrics may be used to characterize people with suicidal ideation~\cite{berman2005forensic}.~\citeauthor{stirman2001word} compare the linguistic expression of poets between a small sample of suicidal and non-suicidal individuals, using a computerized text-analysis program called Linguistic Inquiry and Word Count (LIWC)~\cite{pennebaker2001linguistic}. The authors observed more self-references through first-person singular pronouns, more words related to death, and fewer social references in the poets written by those who did die by suicide~\cite{stirman2001word}. Following that, other studies have used LIWC and similar language analysis techniques to analyze lexical and linguistic features in the text of suicidal individuals from different cultural backgrounds~\cite{fernandez2015linguistic,handelman2007content,lester2010content}. For example, ~\citeauthor{litvinova2017identification}~\cite{litvinova2017identification} use the Russian edition of the LIWC lexicon to analyze the text from blog posts and find that texts written by confirmed suicidal individuals, containing more negation words, fewer social and perception-related words, fewer positive emotion words than texts from a control group.  

All together, these studies provide a core understanding of leveraging mental and psycholinguistic cues for understanding the Papageno effect on social media. Based on the public content shared on the social media platform, we focus on inferring psychosocial outcomes from the perspectives of affect, behavior, and cognition.

\subsection{Social Media and Mental Health Research}



The emergence of social media provides a new powerful ``lens'' to give insights into mental health and suicidal ideation. Prior research uses social media posts to understand more about major depressive disorders~\cite{de2013predicting}, risk suicide behavior~\cite{coppersmith2018natural}, and drug use~\cite{saha2019social}. 


Relatedly, \citeauthor{kumar2015detecting} compare the posting activity and content following celebrity suicides to find a rise in posting frequency and increased suicidal ideation~\cite{kumar2015detecting}. Another work reveals the importance of linguistic features to predict users who move from mental health discourse to suicidal ideation~\cite{de2016discovering}. Based on Reddit postings, the authors develop a propensity score matching to investigate how individuals may discuss their suicidal ideation while controlling for the previous use of linguistic features of mental health. Following this work,~\citeauthor{de2017language}~\cite{de2017language} apply a similar matching framework to study the effect of social support on suicidal ideation risk.
In another work,~\citeauthor{saha2020causal} conduct a causal-inference examination of what factors contribute to improved mental wellbeing in online mental health communities (particularly TalkLife)~\cite{saha2020causal}.

Our work draws motivation from the above body of work in examining the prevalence of the Papageno effect following being exposed to coping story posts on suicidal ideation on social media. Our work adopts the natural language techniques and causal inference analyses to provide a computational framework of measuring this effect and reveals important insights about how people show changes in social media behaviors after engaging with coping story posts.

\section{Data}

Due to the absence of publicly available datasets of coping story posts in social media, we utilize Twitter timeline data of individuals who reply to coping story Twitter posts. The steps of data collection include: 1) collecting Twitter posts, which might describe coping stories;
2) applying coping story classifier
from \cite{metzler2021detecting} and manually verifying the results; 3) collecting timeline data of individuals commenting on the so found coping story posts; 4) building a control dataset from randomly sampled, comparable, individuals. 

\begin{figure}[t]
        \centering\includegraphics[width=0.8\columnwidth]{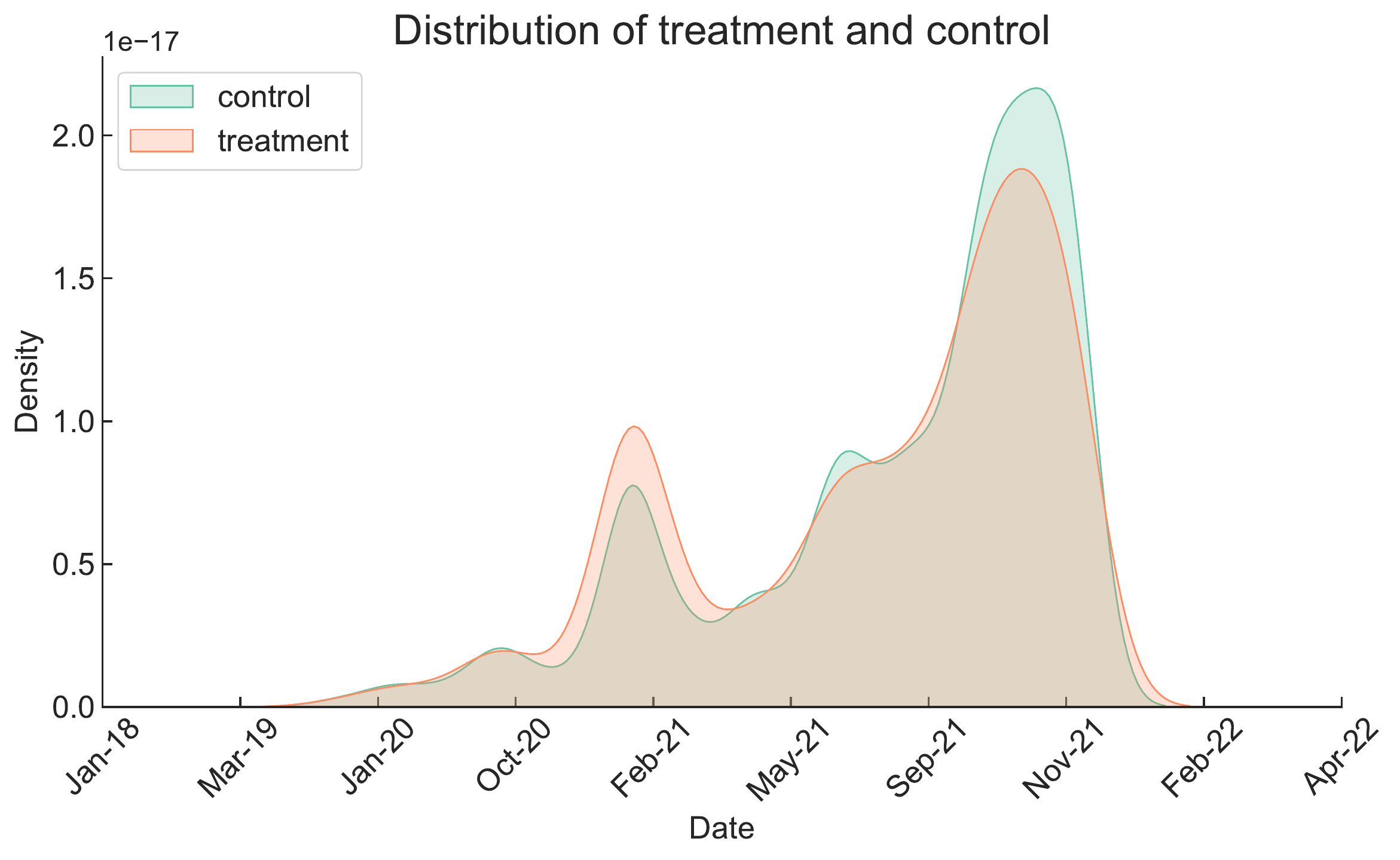}
	\caption{\Tr{} and \Ct{} (placebo) dates distribution.}
        \Description{treatment and control (placebo) dates distribution.}
	\label{fig:treatment_date}
\end{figure}

\subsection{Compiling the Coping Story Posts Dataset} 
The first step is to retrieve Twitter posts that may contain coping stories. This work uses the Twitter Application Programming Interface (API) to collect Twitter posts posted between 1 January 2018 and 1 March 2022. The collected Twitter posts contain at least one term related to suicide attempts, including ``suicidal thoughts,'' ``kill myself,'' ``suicidal ideation,'' and ``end my life,'' and contain at least one of the terms indicating successful coping, including ``happier,'', ``better,'' and ``recover''. After collecting 13,022 Twitter posts, we apply the coping story classifier to annotate each Twitter post as a coping story or a non-coping story. We find 3,077 Twitter posts are annotated as coping story posts. Among them, 709 Twitter posts labeled as coping story posts have at least one reply below them.

\begin{table}[t]
\centering
\sffamily
\small
\caption{Paraphrased example Twitter posts labeled with coping story or non-coping story and Twitter posts responding to coping story posts. \label{tab:sample_tweet}}
\begin{tabular}{p{0.95\linewidth}} 
\toprule
\rowcollight \textbf{\textbf{\textbf{\textbf{\textbf{\textbf{\textbf{\textbf{Coping Story Posts}}}}}}}} \\ 
``I'm posting this because I've had suicide ideas passively for a long time. I finally realized I was suicidal three years ago. I believed that the desire to be better off dead was common. It is NOT the norm. If you have such ideas, you should seek professional assistance.'' \\ 
\hdashline
``When I was a patient in the psychiatric hospital, they had to remove my shoelaces to prevent me from self-injury. Today marks one month without suicide ideation. My life has improved after receiving therapy from all of my physicians. Cheers to the continuation of living in the present!''\\ 
\rowcollight \textbf{\textbf{\textbf{\textbf{Non-Coping Story Posts}}}} \\ 
``Even more terrible than my thoughts of death are my suicide ideas. People told me when I was 10 that it would get better, but it hasn't, and I want to die yet nothing works. It's so unfair that no matter how many times I try, I always fail. I'm sorry if this is frustrating; I just feel so alone.'' \\ 
\hdashline
``But I wanted to kill myself again this weekend. I've never been happier. But every day is so full of grief for the body I don't have and will never be capable of having.'' \\
\rowcollight \textbf{\textbf{\textbf{\textbf{Twitter Posts Responding to Coping Story Posts}}}} \\ 
``Dear friend, I'm happy that things didn't turn out the way you had hoped. I wouldn't want the past few of years to have been any other way because they have been such an adventure. To where we all go in the future is something I am looking forward to see.'' \\ 
\hdashline
``I'm glad to hear that you're doing well. It's good to know that you have support from close friends and family because I am aware of how challenging it may be to handle some circumstances.'' \\
\bottomrule
\end{tabular}
\end{table}

\begin{figure*}[t]
	\centering\includegraphics[width=1.68\columnwidth]{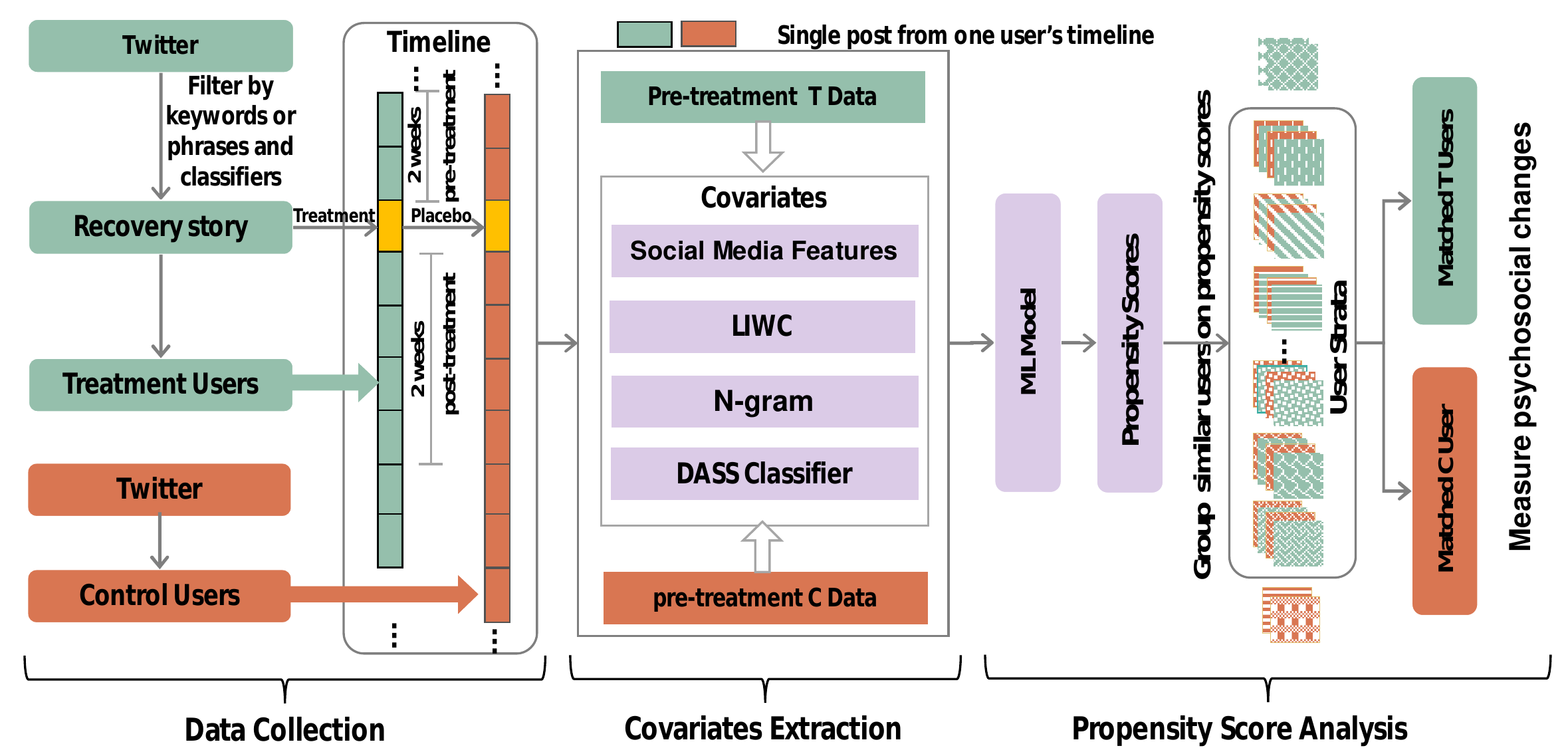}
	\centering
	\caption{Schematic diagram of propensity score matching between \Tr{} individuals and \Ct{} individuals.}
        \Description{Schematic diagram of propensity score matching between Treatment individuals and Control individuals.}
	\label{fig:diagram}
\end{figure*}

\subsubsection{Annotating Coping Story Posts}
We use a multi-label classifier provided by~\citeauthor{metzler2021detecting} to annotate coping story posts in our dataset. It categorizes Twitter posts into the following six categories~\cite{metzler2021detecting}: 
\begin{enumerate}
  \item Personal experiences of coping (coping story): Personal experiences about suicide that express a feeling of hope, healing, methods of coping, or reference alternative options to suicide. The tone could be positive or neutral. Previous studies suggest that such narratives may have a Papageno effect.

  \item Personal experiences of suicidal ideation and attempts: Personal stories about suicide that lack a sense of coping or hope.

  \item Suicide cases: Reports of suicides that have been carried out or prevented

  \item Awareness: Twitter posts raising awareness about suicide, emphasizing high rates or links to problems such as such as bullying, racism, depression, or veteran status.

  \item Prevention: Twitter posts that provide information on solutions or initiatives aimed at addressing the problem of suicide, including prevention at both individual and public health levels.

  \item Twitter posts that do not fall under any of the above categories.

\end{enumerate}

As the aim of this step is to find individuals who comment on coping story posts, we subsequently focus on Twitter posts labeled as a coping story post. In order to verify the ability of the classifier to find Twitter posts that contain actual coping stories, we manually check Twitter posts that are labeled as coping story posts by the classifier.

\subsubsection{Annotation Task}
We randomly sample 400 Twitter posts out of 709 Twitter posts labeled as coping stories that have at least one reply below them. Using the codebook from \cite{metzler2021detecting}, one author independently annotates 400 Twitter posts. If there are any posts that they are unsure about, the author discusses the posts with other authors and together they reach an agreement on how to code them. After finishing the annotation, another author randomly selects 50 posts out of 400 posts to verify the annotation result. We use Cohen's kappa to validate our annotation process. This results in Cohen's $k$ of 0.81 with an agreement of 92.8\%,  which indicates substantial agreement \cite{landis1977measurement}. Out of the 400 posts labeled coping story by the classifier, we find 347 posts are correct predictions, indicating 86.7\% accuracy of the classifier to identify copy story posts.

\subsection{Compiling the Treatment Dataset} 
For the Twitter posts annotated as coping story posts, we assume that the individuals who reply to the Twitter posts might have been impacted by the coping story. For our coping story dataset, we identify 2470 unique individuals who replied to at least one coping story. We collect Twitter metadata, including the number of posts, likes, followers, followees, and the account creation time. To mitigate the confounding effects of engaging with multiple coping story posts, we remove 2 individuals who reply to multiple coping story posts. For the remaining 2468 individuals, we collect their timeline data two weeks before their reply on the coping story and two weeks after. In the end, the target dataset contains 2468 individuals with a total of 787K Twitter posts.

\subsection{Compiling the Control Dataset}
As we seek to isolate the effect of the coping story on individuals, we build a \Ct{} dataset of individuals who do not reply to coping story posts during the investigated period. To do so, a \Ct{} dataset is built with individuals who have similar attributes to the treatment individuals prior to their engagement with a coping story. To find such control individuals, we use keywords, including ``life'', ``job'', ``music'', and ``movie'' to search for individuals on Twitter. For each keyword, we collect 4000 individuals and the timeline of their Twitter posts. For each individual in the \Ct{} dataset, We assign a placebo date from the non-parametric distribution of treatment date in the \Tr{} dataset to any day the \Ct{} individual replies to other Twitter posts to reduce any temporal confounds.
We utilize Kolmogorov-Smirnov (KS) test to measure the similarity in the two distributions (Figure \ref{fig:treatment_date}). The KS test yields a low statistic of 0.05, suggesting that the probability distributions of treatment and placebo dates are similar. In the end, we collect timeline data two weeks before the placebo date and two weeks after of 8465 individuals to build the \textit{control} dataset.

\section{Methods}
\subsection{Study Design and Rationale}

We adopt a causal inference framework~\cite{imbens2015causal} to isolate the Papageno effect. The schematic diagram of our approach is shown in Figure~\ref{fig:diagram}. Our approach first matches individuals of the \Ct{} group with individuals of the \Tr{} group\footnote{\Tr{}: Conventionally, ``\Tr{}'' refers to being given \Tr{} or intervention. 
We use the term ``\Tr{}'' in accordance with the causal inference terminology to differentiate individuals replying to coping story posts and the individuals not replying to coping story posts.} based on pre-\Tr{} behavioral attributes. For this, we train a machine learning classifier to estimate the likelihood of an individual being assigned to either the \Tr{} or \Ct{} group (i.e.,\textit{ propensity}) based on covariates and perform matching across groups using estimated propensity scores. Within matched groups of \Ct{} and \Tr{} groups, we analyze the following psychosocial outcomes between matched \Ct{} groups and \Tr{} groups.
In sum, our approach ensures that members of the \Tr{} group and \Ct{} group who are being compared have similar behavior prior to replying to coping story posts. This gives us the means to analyze the differences inpsychosocial outcomes between the matched \Tr{} individuals and the members of the \Ct{} group.


\subsection{Measuring Psychosocial Outcomes}

To examine the psychosocial effects of engaging with coping story posts on social media, we measure three psychosocial outcomes drawing from psychiatry and psychology literature:  affective, behavioral, and cognitive outcomes~\cite{breckler1984empirical}. We operationalize these measures drawing on prior research in social media and mental health~\cite{saha2018social,saha2020causal}.

\vspace{0.04in} \noindent\textbf{Affective Outcomes}
Affect is defined as any experience of feeling or emotion~\cite{vandenbos2007apa}. As individuals use emotive, relativistic language in their self-motivated online texts, language is an effective way to infer affective psychosocial wellbeing. To estimate affective outcomes, we use the following metrics:

\vspace{0.04in} \noindent\textit{Affective Words.}
We employ the well-validated psycholinguistic lexicon, Linguistic Inquiry and Word Count (LIWC)~\cite{pennebaker2007expressive} to obtain normalized occurrences of words in affective categories per individual. The selection of these measures is inspired by studies like~\cite{ernala2017linguistic, de2013predicting,saha2018social}, where therapeutic symptoms are associated with self-initiated and expressive writing~\cite{cohn2004linguistic,chung2007psychological}.

\vspace{0.04in} \noindent\textit{Symptomatic Mental Health Expressions.} 
Prior research notes the comorbidity of multiple mental health conditions~\cite{rosenblat2016cognitive}, and we operationalize the language indicative of different mental health symptomatic expressions of depression, anxiety, stress, and suicidal ideation~\cite{saha2019social}.
To identify mental health symptomatic expressions in social media language,~\citeauthor{saha2019social}~\cite{saha2019social} develop multiple binary machine learning classifiers based on transfer learning methodologies. These classifiers are $n$-gram–based ($n$=1,2,3) binary support vector machine (SVM) models, and are trained using appropriate Reddit communities (\textit{r/depression} for depression, \textit{r/anxiety} for anxiety, \textit{r/stress} for stress, and \textit{r/SuicideWatch} for suicidal ideation). People in these communities post about mental health symptoms to receive feedback and to support others. The posts in these subreddits are used as training data to identify language used in connection with mental health. The training data for texts not related to mental health originates from non-mental-health-related content on Reddit. These classifiers perform at a high accuracy of approximately 0.90 on average on held-out test data~\cite{saha2019social}, and have also been used in other research~\cite{saha2020causal,saha2020psychosocial}. We use these classifiers to measure the aggregated proportion of expressing mental health concerns per individual. A lower quantity of posts on mental health symptomatic expressions indicates better psychosocial wellbeing.

\vspace{0.04in} \noindent\textbf{Behavioral Outcomes.} Literature in psychology defines behavioral psychological well-being as including three factors: individual's overt actions, behavioral intentions, and verbal statement regarding behavior~\cite{breckler1984empirical}. Previous studies quantify behavioral psychological wellbeing by measuring the shifts in social functioning and interests~\cite{guntuku2019language,saha2018social}. Our work operationalizes the following measures to obtain behavioral outcomes.

\vspace{0.04in} \noindent\textit{Activity.}
We investigate if engaging with coping story posts promotes individuals to be more active on Twitter. Higher activity likely indicates increased extroversion, and is associated with therapeutic benefits~\cite{ernala2017linguistic,saha2018social}. To quantify activity on Twitter, we calculate the average number of Twitter posts per day for every individual.

\vspace{0.04in} \noindent\textit{Interactivity.}
Interactivity is another indicator of an individual showing therapeutic effects~\cite{saha2018social,saha2020causal}. We measure participation in discussions on Twitter as interactivity, indicating social engagement. The metric used is the proportion of replies (to other individuals' posts) per original Twitter post.

\vspace{0.04in} \noindent\textit{Topic Diversity.}
To measure the diversity of topics discussed by an individual, we apply a language model on the posted texts. We capture language semantics by adopting a word embedding model~\cite{mikolov2013distributed}, which represents words in vectors in latent semantic dimensions. In particular, we use 300-dimensional word embeddings pre-trained on Google News.
Then, for each post from the \Tr{} and \Ct{} datasets, we calculate the average cosine distance from the centroid of the corresponding corpus. Higher the average the distance from the centroid, greater is the topical diversity~\cite{wang2021mutual}.

\vspace{0.04in} \noindent\textbf{Cognitive Outcomes.}
The cognitive component of psychosocial health encompasses beliefs, knowledge structures, perceptual responses, and thoughts \cite{breckler1984empirical}. We adopt the following measures to quantify an individual's cognitive behaviors.

\vspace{0.04in} \noindent\textit{Readability.}
Readability measures the complexity of a given text. We employ the Coleman-Liau Index (CLI) to assess the readability per individual. CLI is calculated as, $CLI = 0.0588*L-0.296*S-15.8$, where L is the average number of letters per 100 words and S is the average number of sentences per 100 words. Previous research shows a link between measures of language complexity and long-term improvements in psychosocial wellbeing~\cite{ernala2017linguistic}.

\vspace{0.04in} \noindent\textit{Complexity and Repeatability.} 
Complexity and Repeatability are syntactic measurements that reflect an individual's cognitive state in terms of planning, execution, and memory~\cite{ernala2017linguistic}. We measure repeatability as the normalized count of non-unique words and complexity as the average number of words per sentence. Psychosocial wellbeing positively correlates with language complexity, and negatively with repeatability~\cite{ernala2017linguistic,saha2018social}.

\vspace{0.04in} \noindent\textit{Psycholinguistic Keywords.}
We adopt the LIWC lexicon to analyze the proportion of keywords related to cognition, perception, social context, and linguistic style categories. We consider the following five aggregated categories: (1)  Cognition \& Perception (cause, certain, cognitive, inhibition, discrepancies, tentativeness, perception, see, hear, feel, insight) (2) Social Context (biological processes, achievement, body, family, friends, health, home, humans, money, religion, social, work) (3) Lexical Density \& Awareness (adverbs, article, verbs, auxiliary verbs, conjunctions, inclusive, exclusive, preposition, negation, quantifier, relative) (4) Interpersonal Focus (1st personal pronouns, 2nd personal pronouns, Impersonal pronouns) (5) Temporal References (future, past, present). Prior research highlight the association between these lexicons and cognition~\cite{pennebaker2003psychological}. An increased use of these lexicons is related to better psychological conditions~\cite{cohn2004linguistic,chung2007psychological}.

\subsection{Matching}
\subsubsection{Matching Covariates}
Our goal is to measure the psychosocial outcomes of engaging with a coping story post. Matching is an efficient strategy in Case-Controlled studies to estimate causal effects and minimize the possible occurrence of selection biases. In our case, we build several covariates from the \Tr{} and \Ct{} datasets to control for similar pre-\Tr{} behavior on social media. The first set of covariates comprises Twitter individuals’ social media features (the number of posts, likes, followers, followees, and posting frequency). The second set contains the distribution of word usage in their Twitter timelines. We extract the top 100 unigrams as the second covariates set. The third set includes the psycholinguistic features in their timeline data by measuring the word distribution in the LIWC lexicon. The last set of covariates controls for the average usage of posts related to symptomatic mental health expressions, including depression, anxiety, stress, and suicidal ideation.

\begin{figure}[!t]
\subfloat[\label{sfig:a_matching}]{%
  \includegraphics[height=.45\linewidth,width=.45\linewidth]{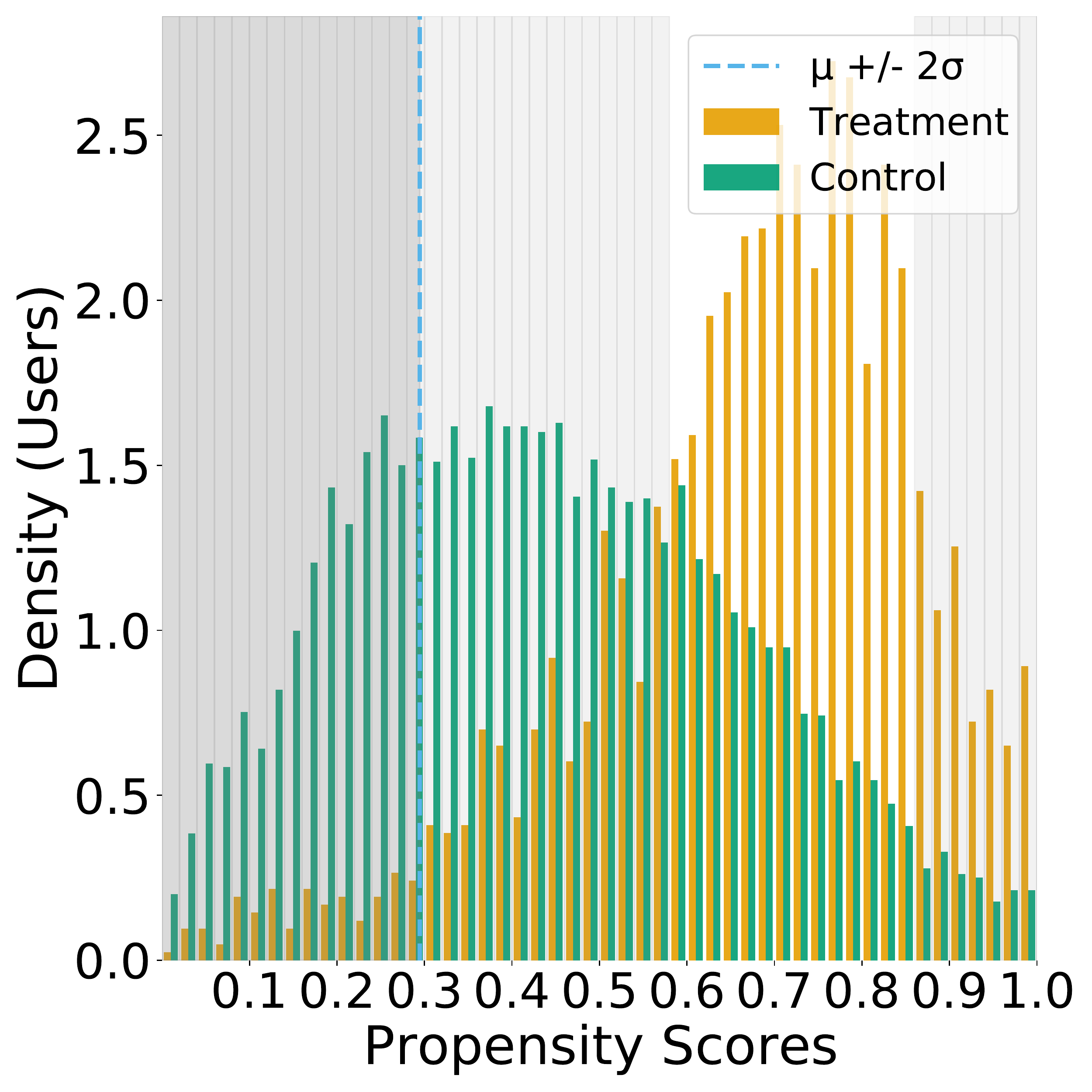}%
}\hfill
\subfloat[\label{sfig:b_matching}]{%
  \includegraphics[height=.5\linewidth,width=.5\linewidth]{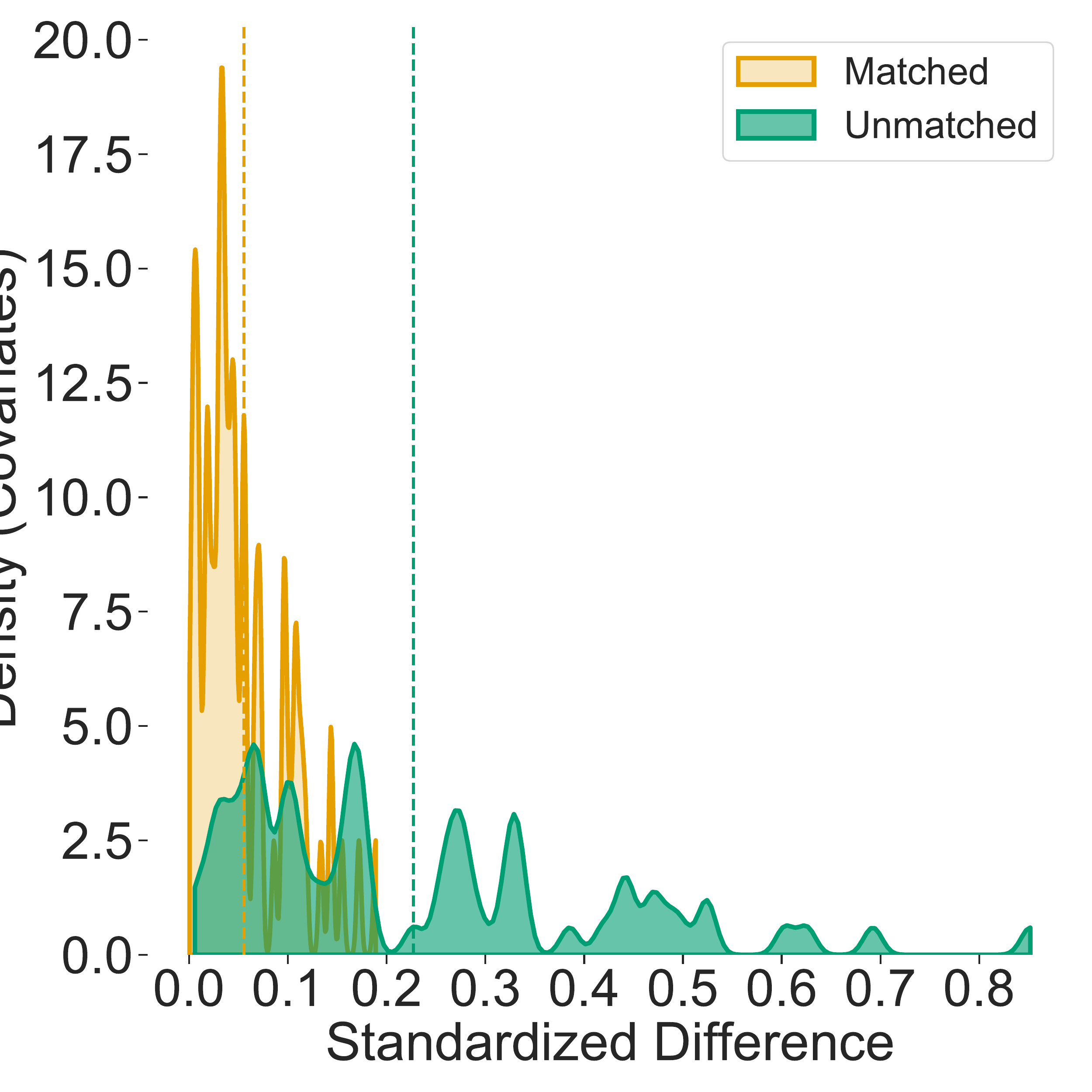}%
}\hfill

\caption{ (a) Propensity score distribution (shaded region are the dropped strata), (b) Quality of matching.}
\Description{ (a) Propensity score distribution (shaded region are the dropped strata), (b) Quality of matching.}
\label{fig:matching}
\end{figure}

\subsubsection{Propensity Score Analysis}
To ensure to have comparable individuals in our \Tr{} dataset and the \Ct{} dataset, matching is employed to pair \Tr{} individuals and \Ct{} individuals, whose covariates are similar to each other. A logistic regression classifier is implement to predict the likelihood of an individual belonging to either the \Tr{} group or \Ct{} group based on their covariates. We divide the propensity score distribution into 50 strata with equal width. The individuals with similar propensity scores are grouped into the same stratum \cite{kiciman2018using}. This helps us to evaluate possible psychosocial outcomes within each stratum, where the \Ct{} group individuals are matched to the \Tr{} individuals based on the pre-\Tr{} behavioral traits. We remove the individuals with propensity scores falling outside two standard deviations from the mean (\autoref{sfig:a_matching} ). We drop the strata failing to satisfy the minimum sample size within each stratum based on previous causal inference research \cite{dechoudhury2016fooddeserts}. By ensuring that there are at least 50 individuals per group in each stratum, this approach results in 14 strata, containing 1,245 \Tr{} and 1,087 \Ct{} individuals.

\subsubsection{Quality of Matching}
To determine whether individuals in the \Tr{} group and \Ct{} group are statistically comparable, we measure the balance of the covariates. We conduct this comparison by calculating the standardized mean differences (SMD) between the two groups in all 14 valid strata~\cite{saha2019social,kiciman2018using}. 
SMD is the difference in the mean covariate values between the two matched groups, divided by the pooled standard deviation.
The two groups can be assumed to be balanced if the SMD of all covariates is lower than 0.2~\cite{kiciman2018using,stuart2010matching,saha2021advertiming}. For the unmatched dataset, the maximum SMD is 0.85, and the mean SMD is 0.22, whereas in our matched dataset, the maximum SMD is 0.19 and the mean SMD is 0.06 (\autoref{sfig:b_matching}). Therefore, this satisfies the threshold of SMD$<$0.2, suggesting that our matching yields balanced \Tr{} and \Ct{} datasets.

\subsubsection{Estimating the Average Treatment Effects}
To estimate the effect of engaging with coping story posts, we compute the relative \Tr{} effect (RTE) for each outcome. For this, we calculate the ratio of likelihood of an outcome measure in the \Tr{} group to that in the \Ct{} group per stratum. Using the number of \Tr{} individuals in each stratum as a weight, we obtain the weighted average RTE per outcome. The outcome is interpreted as an increase (greater than 1) or decrease (less than 1) of observable psychosocial outcomes after engaging with coping story posts compared to the \Ct{} group with similar pre-\Tr{} attributes.

\section{Results}

\begin{figure}[t]
	\centering\includegraphics[width=1\columnwidth]{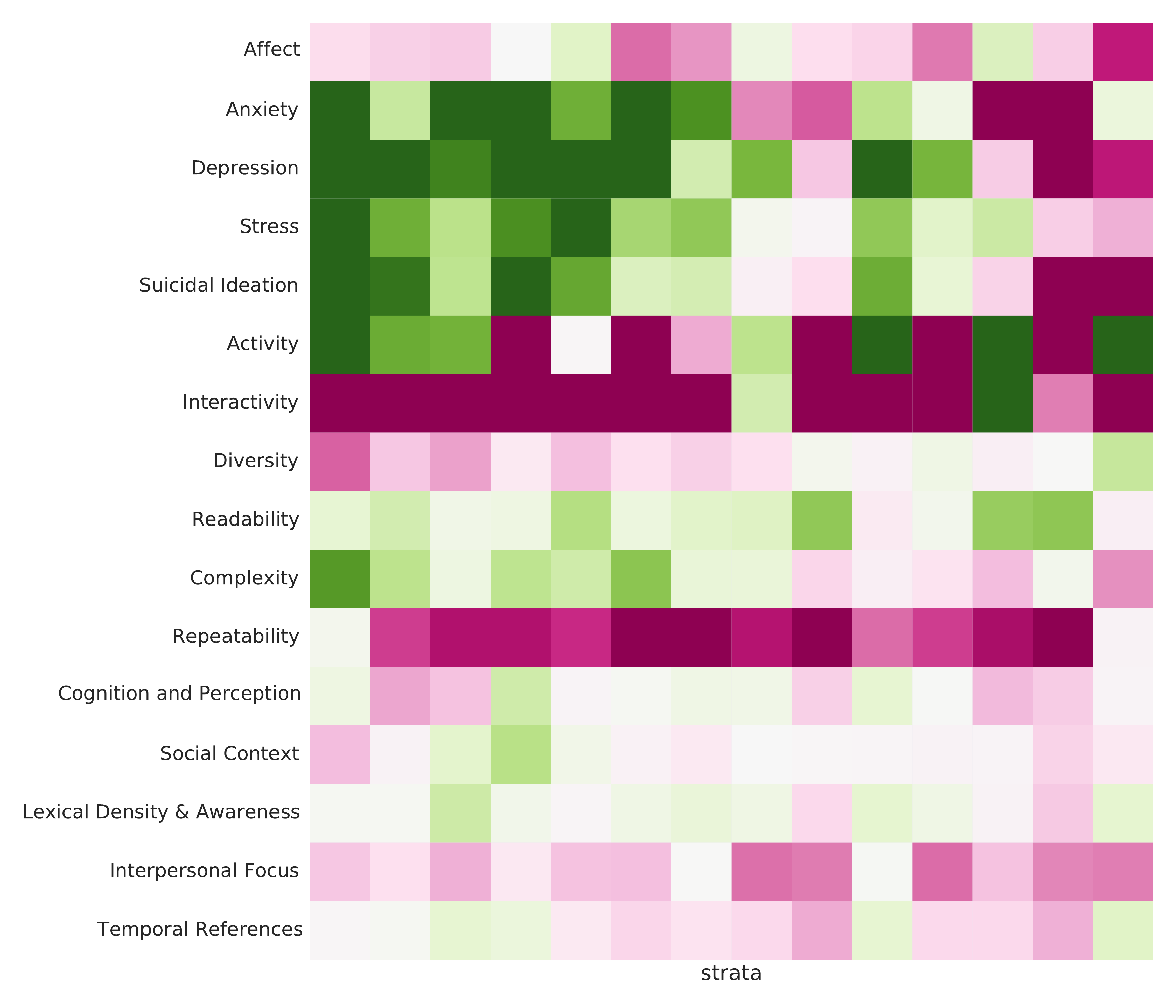}
	\centering
	\caption{RTE per propensity stratum per outcome. The pink color indicates RTE greater than 1, while the green color suggests RTE less than 1.}
        \Description{RTE per propensity stratum per outcome. The pink color indicates RTE greater than 1, while the green color suggests RTE less than 1.}
	\label{fig:heatmap}
\end{figure}

In this section, we present the shifts for each psychosocial outcome across the matched \Tr{} and \Ct{} individuals in the corresponding datasets. We calculate the effect size (Cohen's $d$), and measure statistical significance in differences using an independent sample $t$-test. Figure \ref{fig:heatmap} shows the distribution of RTE per stratum across different psychosocial outcomes. \autoref{tab:result} summarizes these differences. 

\vspace{0.04in} \noindent\textbf{Affective Outcomes.} \textit{Affect.} In \autoref{tab:result}, we observe that individuals in the \Tr{} group use more affective words than the matched \Ct{} individuals after engaging with coping story posts. The average number of affective words used by  \Tr{} individuals is 11\% higher than among \Ct{} individuals. The effect size (Cohen's $d$=0.20) suggests small differences between the two distributions and the t-test indicates statistical significance ($t$=2.19, $p$<0.05). This observation affirms that individuals use more affective words after engaging with coping story posts.

\vspace{0.04in} \noindent\textit{Symptomatic Mental Health Expressions.}
We find that engaging with coping story posts is associated with decreases in the use of symptomatic stress and depression expressions. This is revealed by lower average percentages of symptomatic stress and depression Twitter posts from individuals in the \Tr{} group reflecting stress ($t$=-3.96, $p$<0.001) and depression ($t$=-2.84, $p$<0.01). In contrast, we find no significant differences in our measures of anxiety and suicidal ideation after individuals engage with coping story posts between the two corresponding groups. This illustrates that engaging with coping story posts does not increase the use of symptomatic anxiety and suicidal ideation expressions. These observations are consistent with prior research which indicates that media featuring individuals coping with depression and suicidal ideation reduces depression and shows no effect on suicidal ideation~\cite{niederkrotenthaler2020effects}.

\begin{table}[t!]
\centering
\sffamily
\footnotesize
\caption{Summary of psychosocial differences across all the outcomes between \Tr{} and \Ct{} individuals. We report mean psychosocial outcomes across all matched individuals, effect size (Cohen's $d$), independent sample t-statistic. The p-values from LIWC categories are adjusted using non-negative two stage FDR correction ($ * p <0.05, ** p<0.01, *** p<0.001$).}
\label{tab:result}
\begin{tabular}{llrrrr@{}l@{}}
\setlength{\tabcolsep}{1pt}\\
\textbf{Categories} & \textbf{Tr.} & \textbf{Ct.} & \textbf{RTE} & \textbf{d} & \textbf{t-test} & \\ 
\toprule
\hdashline
\rowcollight \multicolumn{7}{c}{\textbf{Affective Outcomes}}\\
LIWC: Affect & 0.09 & 0.08 & 1.05 & 0.20 & 2.19 &* \\ 
Anxiety & 0.05 & 0.04 & 1.08 & 0.17 & -0.33& \\ 
Depression & 0.15 & 0.17 & 0.93 & 0.28 & -2.84&**  \\ 
Stress & 0.35 & 0.38 & 0.94 & 0.28 & -3.96&*** \\ 
Suicidal Ideation & 0.07 & 0.06 & 1.04 & 0.15 & -0.41& \\ 
\rowcollight \multicolumn{7}{c}{\textbf{Behavioral Outcomes}}\\
Activity & 4.33 & 4.19 & 1.10 & 0.16 & 0.40&  \\ 
Interactivity & 8.89 & 2.78 & 3.34 & 0.34 & 4.17&** \\ 
Topics Diversity &  0.37 & 0.36 & 1.02&  0.25& 2.51&*  \\
\rowcollight \multicolumn{7}{c}{\textbf{Cognitive Outcomes}}\\
Readability & 12.33 & 11.52 & 0.95 & 0.18 & -2.16&* \\ 
Complexity & 9.26 & 9.52 & 0.99 & 0.19 & -1.42& \\ 
Repeatability & 0.51 & 0.45 & 1.11 & 0.30 & 5.09&*** \\ 
LIWC: Cognition \& Perception & 0.27 & 0.27 & 1.02 & 0.17 & 1.12& \\  
LIWC: Social Context & 0.18 & 0.17 & 1.01 & 0.08 & 0.55& \\ 
LIWC: Lexical Density \& Awareness & 0.60 & 0.61 & 0.99 & 0.15 & -1.23& \\ 
LIWC: Interpersonal Focus & 0.12 & 0.11 & 1.08 & 0.26 & 2.77&*  \\ 
LIWC: Temporal Reference &  0.10 & 0.10 & 1.02&  0.17& 1.45& \\ 

\bottomrule
\end{tabular}
\end{table}
\vspace{0.04in} \noindent\textbf{Behavioral Outcomes.} For the second set of outcomes, we find no significant difference in activity after engaging with coping story posts. We find that the average interactivity of the \Tr{} users is higher than the \Ct{}. Both effect size (Cohen’s $d$=0.34) and independent t-test indicate statistical significance ($t$=4.17, $p$<0.01). This might suggest that engaging with coping story posts likely promotes an individual's participation in online discussions. For topical diversity, we measure the diversity of expressed topics in posts after engaging with coping story posts; the effect size informs small differences between \Tr{} and \Ct{} distributions of topical diversity, and the $t$-test affirms statistical significance ($t$=2.51, $p$<0.05). This indicates individuals in \Tr{} group  discuss a broader range of topics after engaging with coping story posts, suggesting higher psychological wellbeing.

\vspace{0.04in} \noindent\textbf{Cognitive outcomes.}
To examine if engaging with coping story posts leads to shifts in cognition, we measure the differences in readability, complexity, repeatability, and psycholinguistic features. Among these, we find no significant differences in complexity, cognition \& perception, lexical density \& awareness, and temporal reference. However, we observe a significant difference in interpersonal focus ($t$=2.77, $p$<0.05). The changes in the usage of pronouns might suggest a shift in how individuals see themselves in relation to others. Although independent sample $t$ indicates a significant difference in readability  ($t$=-2.17, $p$<0.05), the result of cohen's $d$ shows no difference between the two distributions. One unanticipated finding is the \Tr{} individuals show higher repeatability compared to \Ct{}individual ($t$=5.09, $p$<0.001), which suggest lower psychosocial health.

\section{Conclusion and Discussion}
\subsection{Summary of Findings}
In this work, we develop a novel causal inference framework to verify and study the Papageno effect on social media. Using a Twitter dataset with $\sim$2M posts by $\sim$10K individuals, we observe statistically significant psychosocial (affective, behavioral, cognitive) shifts in individuals after engaging with coping story posts. In assessing these psychosocial effects, our causal framework controls behavioral and linguistic covariates across the \Tr{} and \Ct{} groups. We verify that engaging with coping story posts positively impacts individuals' stress and depression, and improves expressive writing, topics diversity, and interactivity. Our results indicate that engaging with coping story posts on social media is associated with positive benefits for one’s psychosocial wellbeing.

\subsection{Implications}
While we do not directly measure shifts in suicidal ideation itself, our findings still provide evidence for the Papageno effect on social media. Our results suggest that engaging with posts describing personal stories featuring coping with suicidal ideation can bring positive impacts on psychosocial wellbeing. Our work provides a methodology to help measure the psychosocial outcomes of the Papageno effect on social media. By focusing on the psychosocial shifts of a large sample of individuals who engage with coping story posts, our work suggests a role for utilizing social media to access the prospective psychosocial outcomes of the Papageno effect.

Our work bears practical implications for preventing suicide. Online communities may develop strategies for the narratives about sharing suicidal behaviors and ideation, allowing vulnerable members in the community more protected. Our work bears design implications for social media platforms in terms of how these platforms can encourage positive and thriving behavior, especially for those struggling with mental health concerns. Platforms such as Reddit, TalkLife, and 7Cups follow community-driven moderation strategies, and these platforms can include in their community norms what kinds of postings and support can help people draw therapeutic benefits. Social media platforms can show more posts with the Papageno effect when individuals search for suicidal information. These may be beneficial for individuals engaging with coping story posts and helpful for individuals with suicidal ideation to seek support and prevent tragic outcomes.

\subsection{Limitations and Future Work}
We acknowledge our study has several limitations, some of which point to intriguing future research directions. We do not account for passive engagement behaviors. For example, we do not know if an individual read a coping story and is affected by it if this individual does not reply to it. Another limitation is the time of our measurement. We only measure the averaged psychosocial outcomes within two weeks after an individual engaged with a coping story post. However, the psychosocial outcomes may vary over time and show fluctuating results~\cite{binmorshed2019mood}.
Future work can explore psychosocial shifts in a fine-grained temporal manner. 

Our study suffers from~\textit{selection biases}. We gather data from those who publicly reply to coping story posts on Twitter, which is likely to be influenced by self-selection bias. We can only collect data from individuals who are active on social media. This is especially true considering the stigma associated with people having suicidal ideation. In a similar manner, the scope of this study solely focuses on Twitter, which might result in incomplete or inaccurate perspectives. Therefore, our observations might not be generalized to other online communities on Twitter or beyond.

We acknowledge the concerns raised by~\citeauthor{king2019propensity}~\cite{king2019propensity}, and are aware of the limitations of the propensity score methods. We chose to use the well-established stratified propensity score method in our study~\cite{saha2020causal,verma2022examining,saha2018social,kiciman2018using}---it offers a more robust approach to mitigate the confounds, by essentially balancing for the bias-variance trade-off. This method, therefore, addresses some of the biases and limitations inherent in propensity score methods~\cite{king2019propensity}.
It is important to note that, despite our best efforts to control for confounding variables, we cannot infer ``\textit{true causality}'' in our study. As the outcomes might have be been influenced by other online and offline factors, e.g., the individual's suicidal behavior history, the length of a coping story post, and the length of the replies. Despite corroboration by a psychiatrist, we cannot be certain based on twitter posts that the individual is personally seriously considering suicide. Therefore the symptomatic variables and outcomes need additional clinical validation. Future work could combine social media data together with clinically validated data which could lead to more generalizable results on the Papageno effect and the role of social media.



\balance{}

\bibliographystyle{ACM-Reference-Format}
\bibliography{references}

\end{document}